\documentclass[prb,twocolumn,showpacs,floats]{revtex4}

\usepackage{graphics} 
\usepackage{epsfig} 
\usepackage{dcolumn}
\usepackage{amsmath}
\usepackage{color}
\usepackage[colorlinks={true}]{hyperref}
\hypersetup{citecolor={blue},filecolor={blue},linkcolor={blue},urlcolor={blue}}

\begin{document} 

\title{Optimal control of universal quantum gates in a double quantum dot }

\author{ Leonardo K. Castelano$^1$, Emanuel F. de Lima$^1$, Justino R. Madureira$^2$, Marcos H. Degani$^3$, and Marcelo Z. Maialle$^3$}

\email{ marcelo.maialle@fca.unicamp.br}

\affiliation{$^1$ Departamento de F\'isica, Universidade Federal de S\~ao Carlos, S\~ao Carlos, SP 13565-905, Brazil}
 
\affiliation{$^2$Universidade Federal de Uberl\^andia, Ituiutaba, MG 38304-402, Brazil}

\affiliation{$^3$Faculdade de Ci\^encias Aplicadas, Universidade Estadual de Campinas, Limeira, SP 13484-350, Brazil}

\date{\today}

\begin{abstract} 

We theoretically investigate electron spin operations driven by applied electric fields 
in a semiconductor double quantum dot (DQD). 
Our model describes a DQD formed in semiconductor nanowire with longitudinal potential modulated by local gating. 
The eigenstates for two electron occupation, including spin-orbit interaction, are calculated and then used to construct 
a model for the charge transport cycle in the DQD taking into account the spatial dependence and spin mixing of states. 
 The dynamics of the system is simulated aiming at implementing protocols 
for qubit operations,  that is, controlled transitions between the singlet and triplet states. 
In order to obtain fast spin manipulation, the dynamics is  
carried out taking advantage of the anticrossings of energy levels introduced 
by the spin-orbit and interdot couplings. The theory of optimal quantum control is  invoked to find the 
specific electric-field driving that performs qubit logical operations. We demonstrate that it is possible to 
perform within high efficiency a universal set of quantum gates 
$\{$CNOT, H$\otimes$I, I$\otimes$H, T$\otimes$I, and T$\otimes$I$\}$, 
where H is the Hadamard gate, T is the $\pi/8$ gate, and I is the identity, even in the 
presence of a fast charge transport cycle and charge noise effects.

\end{abstract} 

%\pacs{xxxxxxxxxxxxxxxxxxxxxxxxxxxxx}

 \maketitle

\section{Introduction}
\label{intro}

There have been numerous proposals for the realization of quantum bits (qubits) in solid state systems aiming 
to host reliable platforms for quantum computation.\cite{supercond-qubit, petta-dots, NV-qubit}
Quantum dots in semiconductors are promising candidates for such platform due to the mature stage
of the technology for semiconductor devices, including advanced growth material processing as well as the ability to integrate 
structures in nanometer scales. Essentially, the quantum dot provides spatial localization for the qubit
 which, in principle, decreases the coupling of the qubit with the neighboring environment, 
 therefore prolonging the quantum coherences. 
There are proposals for quantum dots using either charge\cite{e-charge} or spin\cite{e-spin, e-spin2} 
as qubits, or both,\cite{shi} whereas spins seem to be more 
favorable because of their longer decoherence times due to the weaker nature of the magnetic interactions.
Nonetheless, it is common to have spin decoherence
brought by charge dynamics in several quantum dot systems.~\cite{hugo}
Of course, the coupling with the environment cannot be completely suppressed, as for instance in the case
of the hyperfine interaction of the electron spin with nuclear spins,\cite{koppens}
 or even because the qubits have to interact with the environment to be externally controlled during the qubit initialization, manipulation and readout processes.

Among the proposals for semiconductor quantum dots hosting qubits,\cite{petta-dots} there is
a very controllable architecture of a double quantum dot (DQD)
created by local gates underneath a semiconductor nanowire,\cite{nanowire-dots, petta-prl,nadj} 
which in turn is connected to source and drain
leads [see Fig.~1(a)]. This system is very versatile because the gate voltages that modify
the potential profile along the nanowire can also control the interdot tunneling and can independently control
the electron occupation in each quantum dot. The local gates can also serve as an input for time-dependent electrical fields that, 
via spin-orbit coupling, act as effective magnetic fields on the electron spins.\cite{nadj} 
In essence, this type of nanowire DQDs behave as 
quantum tunneling devices allowing electrical current through themselves
only when the energy levels of the embedded DQD have a favorable alignment. 
This Coulomb blockade behavior produces very distinctive charge
stability diagrams when measuring the current through the DQD.\cite{petta-prl}
  Moreover, the current is allowed only 
when the spin of the electron tunneling the DQD has orientation contrary to that of the spin of the 
unpaired electron already occupying the neighboring dot. This effect is known as the Pauli or spin blockade\cite{spin-block}
 and it is useful to access the spin configuration of the two-electron states via charge current measurements.

A common procedure for manipulating the electron spin in the above mentioned DQDs consists in initializing the system in
spin blockade regime, \textit{i.e.}, assuring that the pair of electrons have parallel spins (triplet state), then applying a gate voltage 
to reinforce the (Coulomb) blockade.\cite{petta-dots} In this double-blockade regime, the spin manipulation 
can be done by oscillating magnetic fields 
(electron spin resonance, ESR) or oscillating electric fields via effective magnetic field of the spin-orbit effect
(electron dipole spin resonance, EDSR). The latter is technologically
more attractive since it is an all-electrical technique.
Finally, the readout is done by lifting the Coulomb blockade and checking if the spin blockade has been 
lifted by the manipulation procedure. If so, the final state would be a two-electron 
state with antiparallel spins (singlet) and a current would flow through the nanowire DQD. 
This scheme implies that the manipulation procedure is carried out in gate voltages detuned from energy level alignments 
(\textit{i.e.}~in the Coulomb blockade regime) where both ESR and EDSR need many oscillations of the applied field to accomplished
the desired spinflip.

There have been, however, alternative approaches utilizing the energy level avoid crossings.\cite{hugo} If the system 
is forced, by a voltage change, through an avoid crossing, 
the Landau-Zener (LZ)\cite{landau, zener} effect can work either in
favor or in opposition to a tunneling between two energy-level avoid branches. The tunneling is favored if the system
is rapidly forced through the avoid crossing. If these avoid crossings are between states
of different spin configurations, for instance due to spin-dependent interactions 
as the spin-orbit interaction, then it is possible to have a faster spinflip process in
the spin manipulation procedure. In fact, the LZ effect was shown to produce much faster 
and stronger spin dynamics than the EDSR excitation, for instance, given rise to strong resonances even 
for excitations with harmonic frequencies.\cite{petta-prl}

In the case of the nanowire DQD discussed above, the LZ tunneling introduces a parallel effect that can
be prejudicial to the spin manipulation. This adverse effect occurs because the nanowire DQD, when operated at the
avoid crossing, populates the singlet state [S(2,0) in Fig.~1(c)] which produces current since this state lifts the
spin blockade and has a strong overlap with the drain lead.
In other words, the LZ process is inevitably accompanied by an unloading process
of the DQD, leaving only one electron localized in the DQD, and allowing for a subsequent reloading by another
electron coming from the source lead [see Fig.~2(a)]. This creates a charge transport cycle through the
DQD in which the loading and unloading processes 
introduce unwanted decoherence channels into the two-electron dynamics. Thus,
from one perspective LZ can speed up the spinflip dynamics, but from another
perspective it can lead to decoherence due to charge dynamics.\cite{hugo,hugo2}
 Our work investigates this situation to understand
to which degree LZ tunneling can be useful in the manipulation of two-electron spins in nanowire DQDs.
 
Our theoretical investigation starts with a model Hamiltonian describing the nanowire DQD
as a quasi-one-dimensional problem of two electrons. Both spatial and spin degrees of freedom
are treated and the Schr\"odinger equation, including spin-orbit interaction and the source-drain applied voltage, 
 is solved for the eigenstates -- in this case for the five lowest energy states: two singlets 
 and three triplets [Fig.~1(c)].
 The eigenstates are used to construct a model for the load and unload of the DQD, where the spatial dependencies and the spin mixing of the eigenstates are taking into account. The load/unload rates
 are dependent on the source-drain applied voltage (called here {\it detuning} for short). There
 is only one free parameter characterizing these rates, which is a global prefactor yielding the 
 intensity of the load/unload process, \textit{i.e.}~it controls the charge cycle frequency in our model.
 The dependence of the (un)load rate on the detuning also means that time-dependent electric fields, applied to
 perform spinflip dynamics, render a time dependence to the rates. This detuning-dependent
 transport cycle is important to address properly the aforementioned question about the
 usefulness of LZ tunneling in the spinflip manipulation carried out close to the avoid crossings.\cite{hugo2}
 
  The manipulation of spins in DQD raises the question of what degree of control one can achieve in performing qubits logical operations. Although transitions between spin states can generally be accomplished by adjusting few parameters of the external field,  it is not appropriate to construct the set of 
universal quantum gates necessary to implement quantum computation.\cite{nielsen}  However, this situation can be tackled by the multi-target formulation of Quantum Optimal Control Theory (QOCT).\cite{multitarget,Kosloff}
 Here, QOCT is invoked to design field control for
 a universal set of quantum gates $\{$CNOT, H$\otimes$I, I$\otimes$H, T$\otimes$I, and T$\otimes$I$\}$, 
where H is the Hadamard gate, T is the $\pi/8$ gate, and I is the identity, in the presence of a fast charge 
transport cycle.
 
 In this paper, Sec.~II describes the two-electron eigenstates, their dynamical occupations and the 
charge transport cycle model. Sec.~III-A
 shows the dynamics of two-electron states in applied pulses of electric field aiming
 to perform some spinflip transitions. In Sec.~III-B, the optimal quantum control is used to
 demonstrate the feasibility of a set of universal quantum gates.
 Sec.~IV contains our final remarks. The Appendix A and B show details of the relaxation rates introduced by the charge transport cycle and of the multi-target QOCT, respectively.

\section{Theoretical framework}

\subsection{Eigenstates}
\label{sec-eigen}

We work within the effective-mass approximation considering two electrons in a nanowire
as in Fig.~1(a).
The Hamiltonian can be written as:\cite{szafram}
\begin{equation}
H=h_1 + h_2 + V_c(| {\bf r}_{2}  - {\bf r}_{1}  | ),
\end{equation}
where $V_c$ is the Coulomb repulsion between electrons. The single-electron Hamiltonians are:
\begin{equation}
h_i = T_i + V({\bf r}_i ) + \frac{1}{2} g(x) \mu_B B \sigma_x 
+H_{SO_i},   ~~(i=1, 2),
\end{equation}
with $T_i$ being the kinetic energy operator,
 $V({\bf r}_i )$  the structure potential, $\frac{1}{2} g(x) \mu_B B \sigma_x $
the Zeeman term for magnetic field along the nanowire ($x$ axis), including a position-dependent effective
g-factor $g(x)$,\cite{petta-prl} and $H_{SO_i}$ the spin-orbit interaction given below.
We assume a strong confinement for the transverse directions of the nanowire, such that the electron motion
can be quantized in these directions and separated from the motion along longitudinal direction.
For the transverse quantized motion, we take\cite{szafram}
the ground state in a cylindrical potential and rewrite the Hamiltonian for the corresponding quasi-one-dimensional
problem along the nanowire as:\cite{szafram}
\begin{eqnarray} 
H &=& \sum_{i=1,2} [ -\frac{\hbar^2 }{2 m^*}\frac{\partial^2 }{\partial x_i^2} + V(x_i) + 
\frac{1}{2} g(x_i) \mu_B B \sigma_{xi} 
+H_{SO_i} ]    \nonumber \\ &&
+ V_c(|x_1 - x_2|),
\label{Hx}
\end{eqnarray}
where $V_c$ is given in Ref.~\onlinecite{szafram-coulomb}, and the orbital effects of the magnetic field are neglected, 
\textit{i.e.}~$T_i=\frac{ ( \hbar {\bf k_i} -A_i/e^2)^2 }{2 m^*} \approx -\frac{\hbar^2 }{2 m^*}\frac{\partial^2 }{\partial x_i^2}$.
 Again, due to the strong transverse confinement, the Rashba
spin-orbit interaction, generated by electrostatic potentials of the applied gates, can be approximated
by:\cite{szafram}
\begin{equation}
H_{SO_i}=\alpha (\sigma_{x_i} k_{y_i}- \sigma_{y_i} k_{x_i})
 \approx -\alpha   \sigma_{y_i} k_{x_i},
\label{HSO}
\end{equation}
where $\alpha$ is the Rashba constant. 

The Schr\"odinger equation is solved for the above Hamiltonian, 
Eqs.~(\ref{Hx}) and (\ref{HSO}), to obtain the energies  $E_n$ and eigenstates:
\begin{equation}
\psi_n(x_1,x_2) = 
\begin{bmatrix}
    \phi_{1n}(x_1,x_2)  \\
    \phi_{2n}(x_1,x_2)  \\
     -\phi_{2n}(x_2,x_1)    \\
    \phi_{3n} (x_1,x_2)    
\end{bmatrix}
  \begin{array}{c}
    | \uparrow   \uparrow \rangle \\
    | \uparrow   \downarrow \rangle \\
     |  \downarrow   \uparrow  \rangle \\
    |  \downarrow   \downarrow  \rangle
\end{array},
\label{spinor}
 \end{equation}
 which are written as spinors in the 1/2-spin basis $\{ | \sigma_{z1} \sigma_{z2} \rangle \} $
 along the $z$ axis. The anti-symmetry by the exchange of the 
two electrons is reinforced by the fact that the spinor components satisfy $\phi_{1,3n}(x_1,x_2)=-\phi_{1,3n}(x_2,x_1)$. 
The method for solving the Schr\"odinger equation
was adapted from a split-operator method\cite{deg-review} to act in the spinor
Eq.~(\ref{spinor}). This method evolves a trial wavefunction in imaginary time,
 resulting in a preferential decay of “high-energy” components of the trial wavefunctions and
is non-unitary. At each simulation step, the
wavefunctions are normalized, orthogonality  between  the  wavefunctions is ensured
using a modified Gram-Schmidt method,\cite{deg-review} and the components $\phi_{1,3n}(x_1,x_2) $
of Eq.~(\ref{spinor}) are anti-symmetrized.   
We used the parameters  $m^*$ = 0.027$m_0$ , $\alpha$ = 110 meV nm, and
the effective g-factor was taken from experiment,\cite{petta-prb}
being $g(x>0)$=6.8 [$g(x<0)$=7.8] for the dot at the right (left) side of the DQD.
Figure 1(b) shows the double-well confinement potential $V(x)$ along the nanowire, 
where the interdot potential barrier was adjusted\cite{comment-triplet20} (35 meV) 
to produce a singlet-triplet splitting of 6 meV as measured in Ref.~\onlinecite{petta-prb}.

%++++++++++++++++++++++++++++++++++++++++++++++++++++++++++++++++++++++
% figure-1 
\begin{figure}[htbp]
\includegraphics[width=9.0cm]{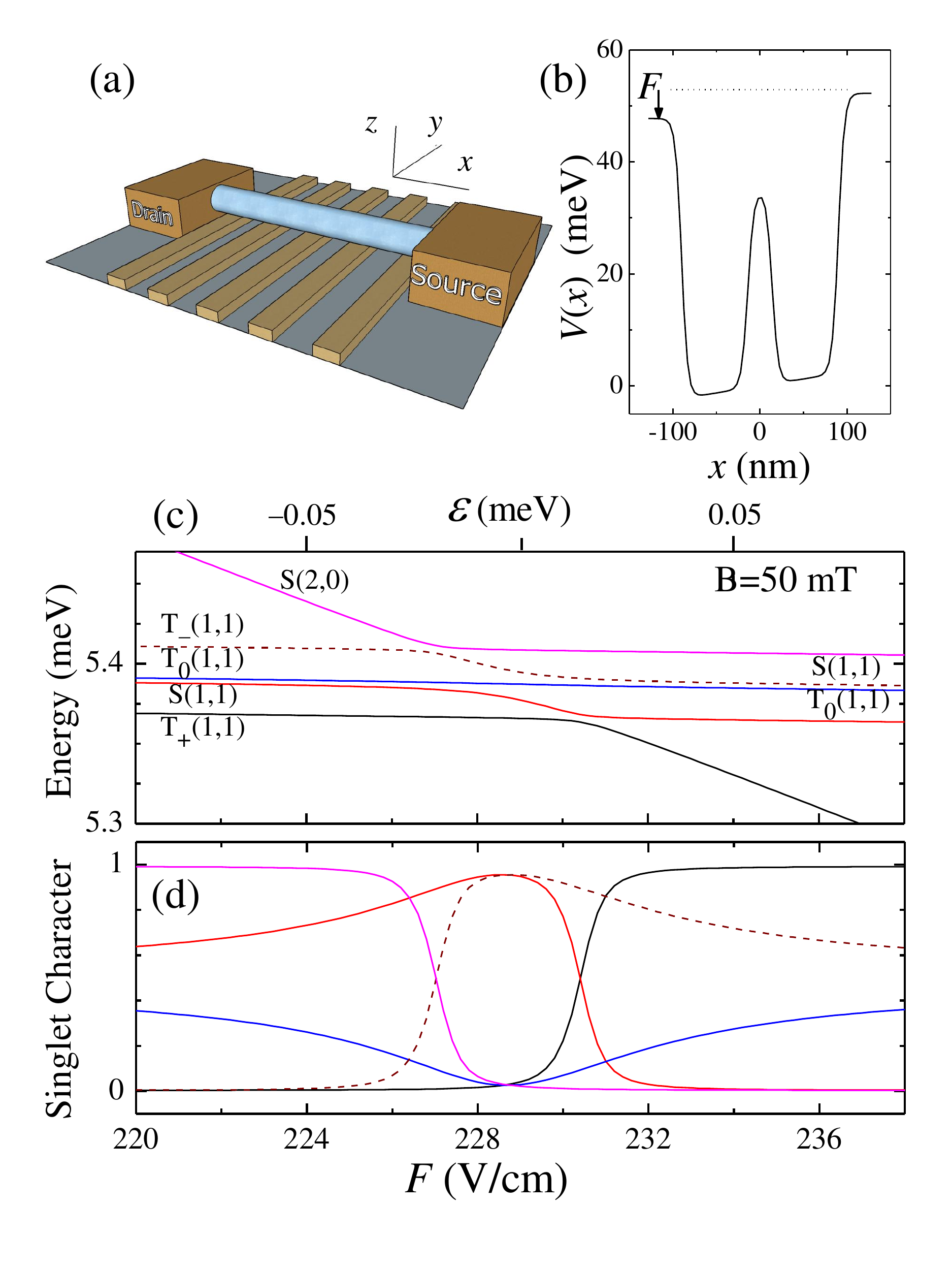} 
\caption{
(a) Schematics of the system investigated. The nanowire is connected
to source and drain leads, and underneath it there are gates to
modulate the electrical potential to create and control the DQD potential profile.
 (b) Potential profile of the DQD along the nanowire used in our calculations.
 The effect of an applied electric field $F$ is shown.
 (c) Energy level diagram for the five lowest energy states as function
 of the source-drain applied electric field $F$ (or detuning $\varepsilon$). 
 We scale $\Delta F$= 100 V/cm to $\Delta \varepsilon$= 1 meV
 for an average system size of 100 nm and $ \varepsilon$= 0 corresponds
 to $ F$= 229 V/cm.
(d) Singlet character of the states shown in (c) as function of the 
detuning. A pure singlet state has character equal 1, whereas for a pure
triplet state the character is zero. Color scheme in (d) is the same as in (c).
}

\end{figure}
%++++++++++++++++++++++++++++++++++++++++++++++++++++++++++++++++++++++++

Panel (c) of Fig.~1 shows the energies of the five lowest energy states as function of the
detuning energy, $ \varepsilon= - e (x_1+x_2)F$, due to an electric field $F$ applied along the wire. 
All results in this paper are for an applied
magnetic field of $B$=50 mT. From now on, the detuning energy $ \varepsilon$ will
be used taking as reference the applied field at the anticrossing, \textit{i.e.}~ zero detuning ($\varepsilon$=0)
means applied electric field of $F$=229 V/cm. 
It is noticed in Fig.~1(c) that only one state
has a pronounced dependence on the detuning. This state has a strong singlet character with two electrons
on the left dot of the DQD: we call it S(2,0). The other four states, with weaker dependence on the detuning,
have one electron in each dot: singlet S(1,1) and triplets T$_+$(1,1), T$_0$(1,1) and T$_{-}$(1,1), where
the indices $\pm, 0$  give the total spin component of the state, and $(n_L,n_R)$ gives the
left and right occupations of the DQD. 
As the state S(2,0) approaches the (1,1) states, some avoid crossings occur. The larger anticrossing at zero detuning,
between the singlets S(2,0) and S(1,1), is due to the interdot coupling and is spin independent.
The smaller anticrossings between S(2,0) and T$_\pm$(1,1) result from the spin-orbit interaction. This
is the spin-mixing term which allows the spinflip dynamics we are interested in. 
We notice that similar spin-mixing can be introduced by electron-nuclei spin interactions,\cite{koppens} 
but we are not including this in our calculations.

Not shown in the diagram of Fig.~1(c) is the triplet manyfold T(2,0) which is,
 as mentioned, $\sim$6 meV above in energy.\cite{comment-triplet20}
The T$_0$(1,1) level 
splits from the S(1,1) away from the anticrossing region, and it wiggles around zero detuning. These effects
result from the position-dependent g-factor $g(x)$.
It is important to stress the fact that all the states have mixed singlet-triplet characters, and this is a function of the detuning.
In Fig.~1(d) we plot the singlet character of the states, defined by the degree of exchange between the particles 1 and 2.
It is seen that the state S(2,0) is mostly singlet only for detunings away from zero.
However, as mentioned, the position-dependent g-factor mixes the singlet state S(1,1) with T$_0$(1,1) for
a broader range of detuning values, whereas the spin-orbit interaction mixes the singlet-triplet
states as observed for the states S(2,0) and T$_\pm$(1,1) in the avoid crossing regions.

\subsection{Dynamics under time-dependent detuning fields}

In this section, we describe the time evolution of the two-electron system
when subjected to time-dependent detunings.
Consider that initially ($t$=0) the system is in a given states $| \psi (t=0) \rangle$, for instance,
one of the eigenstates of Fig.~1(c) for a given detuning $\varepsilon_0$.
If the detuning is modified, say assuming values $\varepsilon (t)$, the initial states
will evolve as a mixing of the complete set of eigenstates. The dynamics
can be calculated by the master equation for the density matrix $\rho (t)= | \psi(t) \rangle 
\langle \psi (t) |$:
\begin{equation}
\frac{ \partial \rho}{\partial t} = \frac{1}{i \hbar} [ H(t), \rho ]+\mathcal{D} \left[\rho\right],
\label{master}
\end{equation}
with $H(t)$ being given by Eq.~(\ref{Hx}) with the inclusion of the time-dependent
applied electric field $F(t)$ along the nanowire: 
$V(x_i) \rightarrow V(x_i)  - e x_i F(t)$, where $e$ is the electron charge. The last rhs term in Eq.~(\ref{master}) is the dissipator that takes into account incoherent effects as described bellow.

Equation (\ref{master}) can be projected onto a set of eigenstates using 
$| \psi (t) \rangle = \sum_n a_n (t) | \psi_n \rangle$,
\begin{equation}
\frac{ \partial \rho_{nm}}{\partial t} = \frac{1}{i \hbar} \sum_l ( H_{nl} \rho_{lm} - \rho_{nl} H_{lm} ).
\label{master2}
\end{equation}
The sum should run over a complete set of eigenstates, but in our case we approximated it by the 
five states calculated around zero detuning as given in Fig.~1(c). In the range 
around $\varepsilon = 0$, this approximation proved to be good.\cite{comment-triplet20}
We call $\varepsilon_0$ the detuning used to project Eq.~(\ref{master}) in
 the corresponding set of states $\{ | \psi_n \rangle \}$.
Incoherent effects are included in Eq.~(\ref{master2}) within the relaxation-time approximation as
transition rates between different states in our vector space, and are given by the Lindblad operators 
$L[A] \rho$ [see Appendix A]. 
While Eq.~(\ref{master2}) is projected onto the reference set $\{ | \psi_n \rangle \}$ 
calculated at $\varepsilon_0$, the incoherent states transitions have to be defined in terms of the instantaneous
 set $\{ | \psi_\alpha \rangle \}$ calculated at the instantaneous detuning $\varepsilon (t)$.
We use for the latter states Greek letter indices.
This allows for detuning-dependent transition rates, which better describe the incoherent dynamics when 
spin-mixing is important, as in the situations closer to the avoid crossings. 
 The incoherent contribution to the dynamics is written as additional terms
to Eq.~(\ref{master2}), and reads:

\begin{equation}
\mathcal{D} \left[\rho_{\alpha\beta}\right]=
\sum_{\gamma,\delta}  \Gamma_{\alpha\beta,\gamma\delta} \enspace \rho_{\gamma\delta},
\label{master-incoh}
\end{equation}
where $\Gamma_{\alpha\beta,\gamma\delta}$ are defined for transitions between 
eigenstates calculated at the instantaneous detuning $\varepsilon (t)$.
Before adding Eq.~(\ref{master-incoh}) to Eq.~(\ref{master2}), we must change basis
$\{ | \psi_\alpha \rangle \} \rightarrow \{ | \psi_n \rangle \}$, yielding
\begin{equation}
\mathcal{D} \left[\rho_{nm}\right]=
\sum_{p,k}  M_{nm,pk} \enspace \rho_{pk},
\label{master-incoh2}
\end{equation}
\begin{equation}
M_{nm,pk}=\sum_{\alpha,\beta,\gamma,\delta} 
\langle \psi_n | \psi_\alpha \rangle     \langle \psi_m| \psi_\beta \rangle 
  \Gamma_{\alpha\beta,\gamma\delta} 
  \langle \psi_\gamma | \psi_p \rangle     \langle \psi_\delta| \psi_k \rangle .
\label{m-incoh}
\end{equation}
The transition rates $\Gamma_{\alpha\beta,\gamma\delta}$ are discussed in the next
subsection and they are explicitly given in the Appendix A.

\subsection{Charge transport cycle}

Nanowire DQD systems work as tunneling devices. The system has a charge transport cycle
that loads and empties the DQD. The specific charge cycle we are investigating
starts with only one electron in the left quantum dot, \textit{i.e.}~occupation (1,0)
as shown in Fig.~2(a). Then, 
a second electron is loaded to the right quantum dot [closer to the source lead, cf.~Fig.~1(a)],
creating a state with occupation (1,1). This state can be either singlet, S(1,1), or triplet, T$_0$(1,1),
T$_\pm$(1,1). The singlet has no restriction imposed by the spin blockade and
it can, if energy level alignment favors, couple to the singlet S(2,0). This state,
with two electrons on the left quantum dot, being closer to the drain lead,
produces current and empties the DQD, returning the system to the initial occupation (1,0).
On the contrary, the triplet states (1,1) are spin blocked, which allows for
an initialization procedure for the two-electron system in a
known spin configuration.

In this work, we construct a model for the charge transport cycle representing the load and unload processes
described above. For that, we introduce an auxiliary state 
representing the one-electron state $| (1,0) \rangle $.\cite{petta-prb}
This state is added to the five two-electron eigenstates already discussed. 
As mentioned above, the load/unload of the DQD is governed by singlet and triplet characters
of the states. However, as shown in Fig.~1(d), the spin characters are functions of the
detuning and, in addition, the dynamics is intended to be done under time-dependent detunings.
Because of this singlet-triplet mixing, we have to ascribe rates for the load and unload processes 
connecting $| (1,0) \rangle $ to all the other five two-electron states. This is represented
in Fig.~2(b), where $\gamma_{L(U)}$ is a global prefactor that is used to control the
intensity of the processes and they are the only free parameters in the model. The
detuning dependent rates for each state $\{ | \psi_\alpha \rangle \}$ are given
by $l_\alpha$ and $u_\alpha$ as discussed next.

The eigenstates calculated in Sec.~\ref{sec-eigen} are used to obtain the detuning
dependent load/unload rates $l_\alpha$/$u_\alpha$. 
The proximity of the two-electron state with the source and drain leads
is also important, so we show in Figs.~2(c) and 2(d) the probabilities of finding one electron on
the left and right quantum dot of the DQD, respectively.
We create a rate for the loading process $l_\alpha$ to be proportional to the 
probability of the state to be closer to the source lead at the right, Fig.~2(d). 
Now, the probability for unloading $u_\alpha$ is given by the product of the probabilities of the state to be closer
to the drain [Fig.~2(c)] times the singlet character of the state, given in Fig.~1(d). The latter
is to ensure the lift of the spin blockade. The resulting unloading rates $u_\alpha$ 
as function of detuning are in Fig.~2(e).

%++++++++++++++++++++++++++++++++++++++++++++++++++++++++++++++++++++++
% figure-2 
\begin{figure}[htbp]
\includegraphics[width=9.0cm]{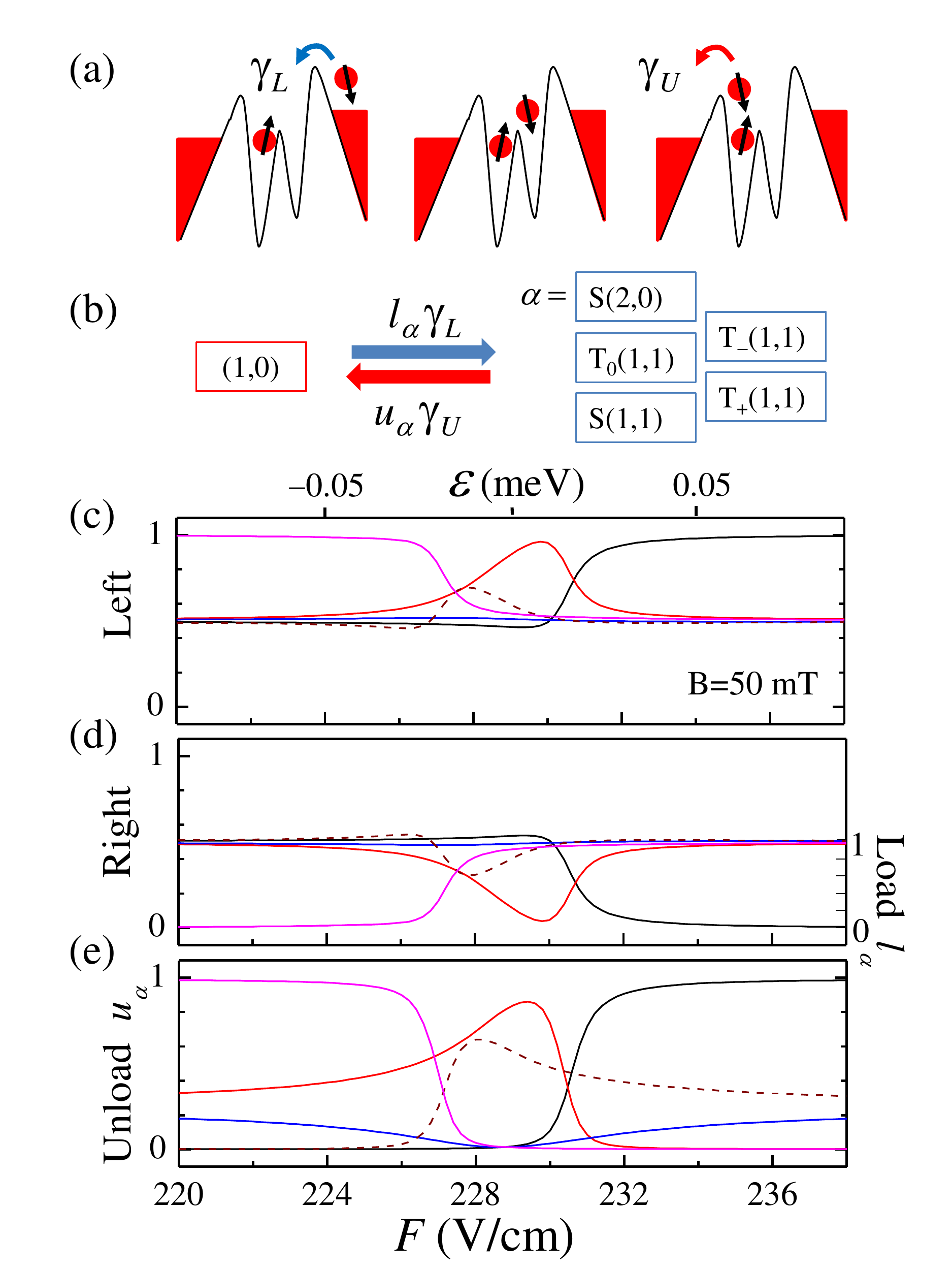} 
\caption{
(a) Schematics of the electron transfer representing the load and unload of the DQD. 
(b) Flowchart of the load and unload processes through the auxiliary state $| (1,0) \rangle $.
Probability for finding an electron, for each eigenstate,
 in the (c) left and (d) right quantum dot of the DQD as function of the detuning. 
(d) equally shows the renormalized rates for loading the DQD,  $l_\alpha$. 
(e) Unloading rates $u_\alpha$ as function of the detuning.
In (c)-(e), the color scheme is the same as in Fig.~1(c).
} 

\end{figure}
%++++++++++++++++++++++++++++++++++++++++++++++++++++++++++++++++++++++++

Finally, the rates $l_\alpha$ and $u_\alpha$ are used to obtain $\Gamma_{\alpha\beta,\gamma\delta}$
in Eq.~(\ref{master-incoh}) in the instantaneous basis set $\{ | \psi_\alpha \rangle \}$.
We use Lindblad superoperators to take into account the incoherent contributions and the details of such a derivation is described in the
Appendix A.

\subsection{Detuning-dependent rates: numerical procedure }

We briefly describe the numerical procedure used to include the detuning-dependent rates
into the time evolution Eqs.~(\ref{master2})$-$(\ref{master-incoh}).

The system is initialized in a given detuning, which can be the one used as reference to
project Eq.~(\ref{master2}), \textit{i.e.}~$\varepsilon_0$. The initial state is chosen and written in terms of 
a density matrix with components projected in the reference basis $\{ | \psi_{n'} \rangle \}$.
When the detuning changes in time, the dynamics follows Eqs.~(\ref{master2})$-$(\ref{master-incoh}),
and the incoherent processes of load/unload of the DQD are calculated at the instantaneous
basis set $\{ | \psi_\alpha \rangle \}$, as discussed in the Appendix A. For each change
in the detuning, a transformation between basis sets
 $\{ | \psi_\alpha \rangle \} \leftrightarrow \{ | \psi_n \rangle \}$ is needed, as
shown in Eq.~(\ref{m-incoh}). Numerically, in order to speed up the calculations,
a number of basis sets for given values of detunings are previously calculated, and so are the matrices  Eq.~(\ref{m-incoh}).
As the detuning varies in time, we interpolate the instantaneous detuning to the closest one previously calculated.
We have calculated a set of 150 basis sets, spanning from detuning fields $F$=214$-$244 V/cm, in step of $\Delta F$=0.2 V/cm.
This range of $F$ comprises the avoid crossings [cf.~Fig.~1(a)], where the dependence on the detuning
is mostly important. It is also the range of detunings we perform the spinflip dynamics
under the LZ effects in this work.

\subsection{Multi-Target Optimal Control}

QOCT is often concerned with driving an initial known state to a desired target state by means of the shaping of an external control field.\cite{qoct} There are several control algorithms to perform this goal. In particular, the control field can be efficiently designed through a monotonically convergent algorithm known as two-point boundary-value quantum control paradigm, where the optimized control field is found interactively.\cite{tbqcp} Of extreme relevance for quantum computing is the fact that we can formulate the QOCT problem to optimize several transitions simultaneously with the same external field. This kind of optimization is closely related to the implementation of a quantum gate, which is a unitary transformation that acts on any linear combination of the logical basis states.
 The key is to construct an optimized field that acts appropriately on each 
 state of the logical basis plus a particular linear combination of all states of the logical basis. This last constraint imposed to the optimized field is necessary to avoid relative phase errors.\cite{multitarget,Kosloff}
 We refer to this approach as the multi-target optimal control algorithm 
 and we apply this procedure to our system in order to find optimized electric fields $F(t)$ 
 aiming at performing the set of universal quantum gates.   Details of the multi-target QOCT used here are in the Appendix B.

\section{Results and discussion}

In this section we present the simulation results and discuss
the effects of the incoherent processes due to the charge transport cycle.

\subsection{Spin dynamics under charge transport cycle}

To illustrate the control of the spin dynamics in the nanowire DQD,  we present results for
 applied detuning pulses which force the system through an energy avoid crossing. In Fig.~3(a) we demonstrate the charge cycle effects, first without the
detuning pulses. The system is prepared in the initial
state $| (1,0) \rangle $ and the detuning remains fixed at 
$\varepsilon_0$=$-0.09$ meV [or $F$=220 V/cm, cf.~Fig.~1(a)].
The charge cycle intensity is set as $\gamma_U=\gamma_L$= 2 GHz.\cite{petta-prl}
The evolution shows the initial state being distributed among the other two-electron states,
and a stationary regime sets in with the occupation of the triplet states T$_\pm$(1,1) being $\sim 50\%$ each,
all the other being close to zero. The evolution to this stationary state
has been mentioned before as the initialization process under spin blockade, 
in this case without the thermalization effects,
\textit{i.e.}~processes that favor transitions from the more energetic state to the less energetic states.
The time necessary to reach the stationary regime and the steady-state values of the triplets T$_\pm$(1,1)
 depends on the $\varepsilon_0$, as shown in Fig.~3(c), where
we note a strong $\varepsilon_0$ dependence of the occupations around the zero detuning.
The reason for the unbalance between steady values of T$_\pm$(1,1) has to do
with the load/unload rates which are detuning dependent. In the case $\varepsilon_0 \lesssim 0$, T$_-$(1,1)
mixes more with S(2,0) than T$_+$(1,1) does, consequently for T$_-$(1,1) the unload rate is stronger
than its (re)load rate, therefore favoring a larger steady occupation of T$_+$(1,1).
This is consistent with the dependence of the rates on the detuning as given in Figs.~2(d) and 2(e). 
Figure 3(b) shows the occupation of the state
$| (1,0) \rangle $ which gives a measure of the current through the DQD, 
$I \sim \gamma_L \sum_\alpha l_\alpha  \rho_{66}$, where  the 
index 6 refers to the $| (1,0) \rangle $ state. The current is
more intense around zero detuning, meaning that the Coulomb blockade is lifted at $\varepsilon_0 \simeq 0$.

%++++++++++++++++++++++++++++++++++++++++++++++++++++++++++++++++++++++
% figure-3 
\begin{figure}[htbp]
\includegraphics[width=9.0cm]{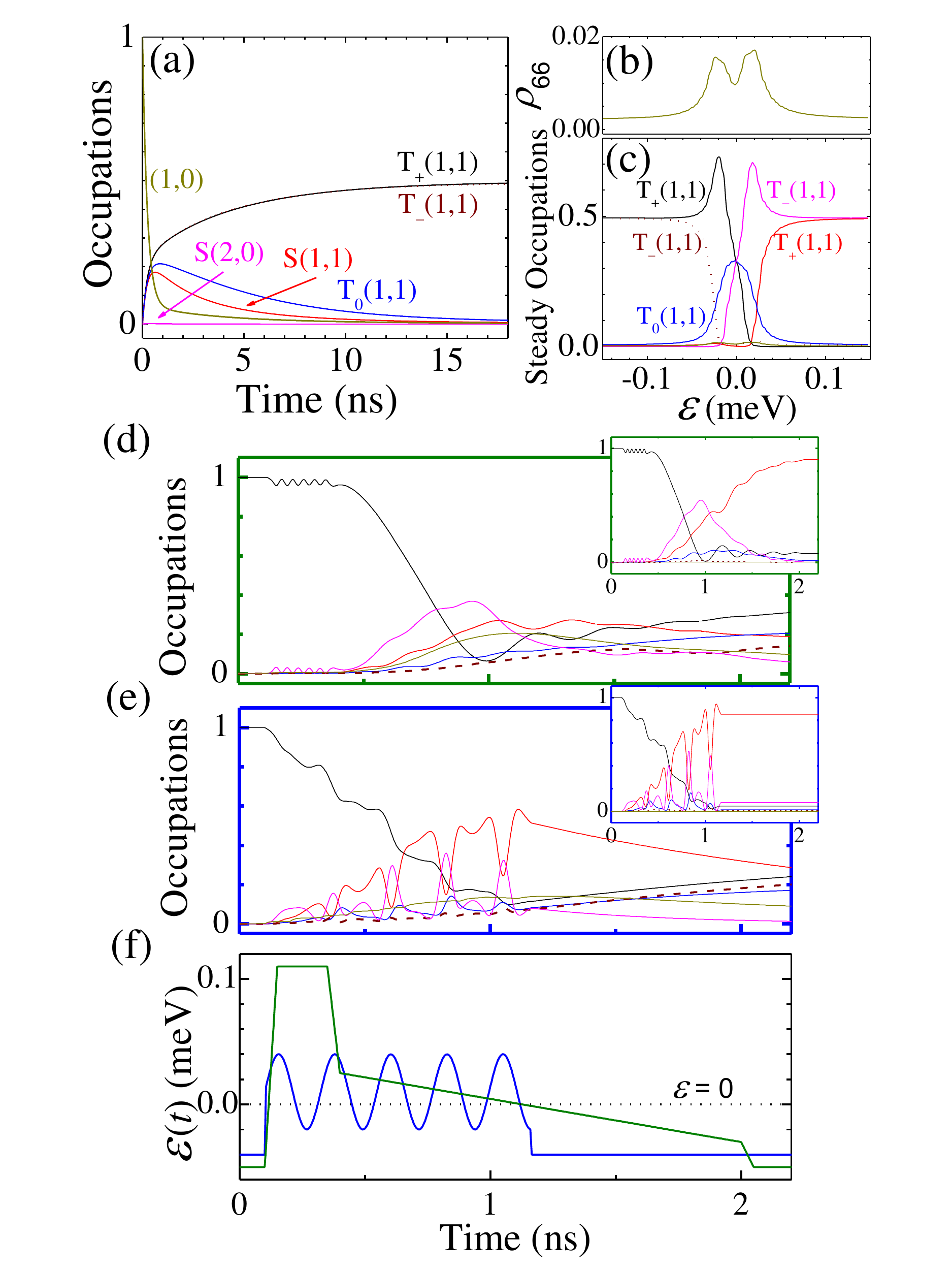} 
\caption{   (a, b, c) The initial state is  $| (1,0) \rangle $ and
the detuning is kept constant at $\varepsilon_0$=$-$0.09 meV. Charge cycle rates are 
$\gamma_{U(L)}$= 2 GHz. 
(a) State occupations as function of time. 
(b)  $| (1,0) \rangle $ occupation and
(c) the other state occupations after 30 ns of simulation.
(d) Simulation with a stepped detuning pulse given in (f) by the green curve. 
The state projections are in the eigenstate basis calculated at $\varepsilon_0$. 
The color scheme is the same as in (a).
(e) Simulation with a sinusoidal pulse given in (f) by the blue curve. 
Insets in (d) and (e) show similar dynamics but with zero charge cycle $\gamma_{U(L)}$= 0.
(f) Detuning pulse profiles as function of time.   }

\end{figure}
%++++++++++++++++++++++++++++++++++++++++++++++++++++++++++++++++++++++++

Subsequently, the initial state is driven through an avoid crossing by a detuning pulse.
The initial state is now chosen to be the triplet T$_+$(1,1) at $\varepsilon_0$=$-$0.05 meV.
We exemplify an attempt to make the transition T$_+$(1,1) $\rightarrow$ S(1,1),
starting and finishing at the same detuning $\varepsilon_0$.  
The pulse profile  [Fig.~3(f), green color] has a forward ramp 
(increasing $\varepsilon$) that is faster than the return ramp.
This choice is intended to control the LZ tunneling probability which
depends on the speed of approaching the avoid crossing. 
The initial fast approaching to the avoid crossing T$_+$(1,1)$-$S(2,0) 
favors the state to remain T$_+$(1,1), and the slow return 
favors the transition to S(1,1).
Figure 3(d) shows the occupations $\rho_{n'n'}$ in time as seen by projections of the state
on the eigenstate basis at $\varepsilon_0$, $\{ | \psi_{n'} \rangle \}$. 
As expected, the forward passage through the avoid crossing kept mostly the state at
T$_+$(1,1), and the slow return enhanced the transition to the
target state S(1,1) with a efficiency of $\sim 20\%$ right at the end of the
pulse ($t$=2 ns). Applying the same scheme, but with zero charge cycle rates $\gamma_{U(L)}$= 0, 
the inset in Fig.~3(d) show a much better efficiency of $\sim 90\%$.

The above example of spin dynamics with detuning pulses operating at the avoid crossings
shows an important aspect of the nanowire DQD system which we have already mentioned
in Sec.~\ref{intro}. The spin dynamics is mediated by the state S(2,0), which is also
the state that triggers the charge cycle and its corresponding incoherent dynamics.
The slow return ramp in the pulse profile of the above example, needed to
enhance the LZ transition T$_+$(1,1) $\rightarrow$ S(1,1), populates S(2,0).
Moreover, the slow speed of this ramp makes things worse because the system
stays in strong charge cycle regime while performing the desired transition.
As a result, we obtained a small efficiency for the transfer T$_+$(1,1) $\rightarrow$ S(1,1)
when including charge cycle relaxation.

There are however means to overcome this prejudicial aspect controlling
several parameters that defines the dynamics of the LZ transitions.
These parameters alter the phase differences between
the state components when the state goes through an avoid crossing,
and therefore they affect significantly the interference effects behind the LZ transition. 
Similarly, there is the possibility of using multiple passages through the avoid crossing.
This is interesting because it can have a cumulative effect. Consequently, faster detuning ramps
can be repeatedly used to accomplish a desired transition, however
without the nanowire DQD being subjected to the strong charge cycle for long period of times.
In Fig.~3(f), the blue curve depicts a pulse profile of another attempt to perform 
the transition T$_+$(1,1) $\rightarrow$ S(1,1),
starting and finishing at the same detuning $\varepsilon_0$. It consists
of an initial very fast drive from $\varepsilon_0$=$-$0.04 meV to $\varepsilon_c$=0.01 meV, through the
avoid crossing T$_+$(1,1)$-$S(2,0), followed by five back and forth  avoid crossing passages in
a sinusoidal form, $\varepsilon(t)=\varepsilon_c+\varepsilon_{ac} \sin (2\pi f_0 t)$, 
where $\varepsilon_{ac}$= 0.03 meV and frequency ($f_0$=4.45 GHz) matching the 
energy of the transition T$_+$(1,1) $\rightarrow$ S(1,1) at $\varepsilon_0$.
Finally, the detuning is returned to $\varepsilon_0$.
In Fig.~3(e), we plot the state occupations as function of time for
projections onto the eigenstate basis for  $\varepsilon_0$.
The cumulative effect is clearly seen as the initial state T$_+$(1,1)
is progressively transferred to the target state S(1,1). In this case,
the transfer efficiency is $\sim 52\%$ (at $t$=1.2 ns) for the charge cycle rates
$\gamma_{U(L)}$= 2 GHz, and $\sim 86\%$ (inset)
without charge cycling, \textit{i.e.}~$\gamma_{U(L)}$= 0.

 %++++++++++++++++++++++++++++++++++++++++++++++++++++++++++++++++++++++
% figure-4 
\begin{figure}[htbp]
\includegraphics[width=9.0cm]{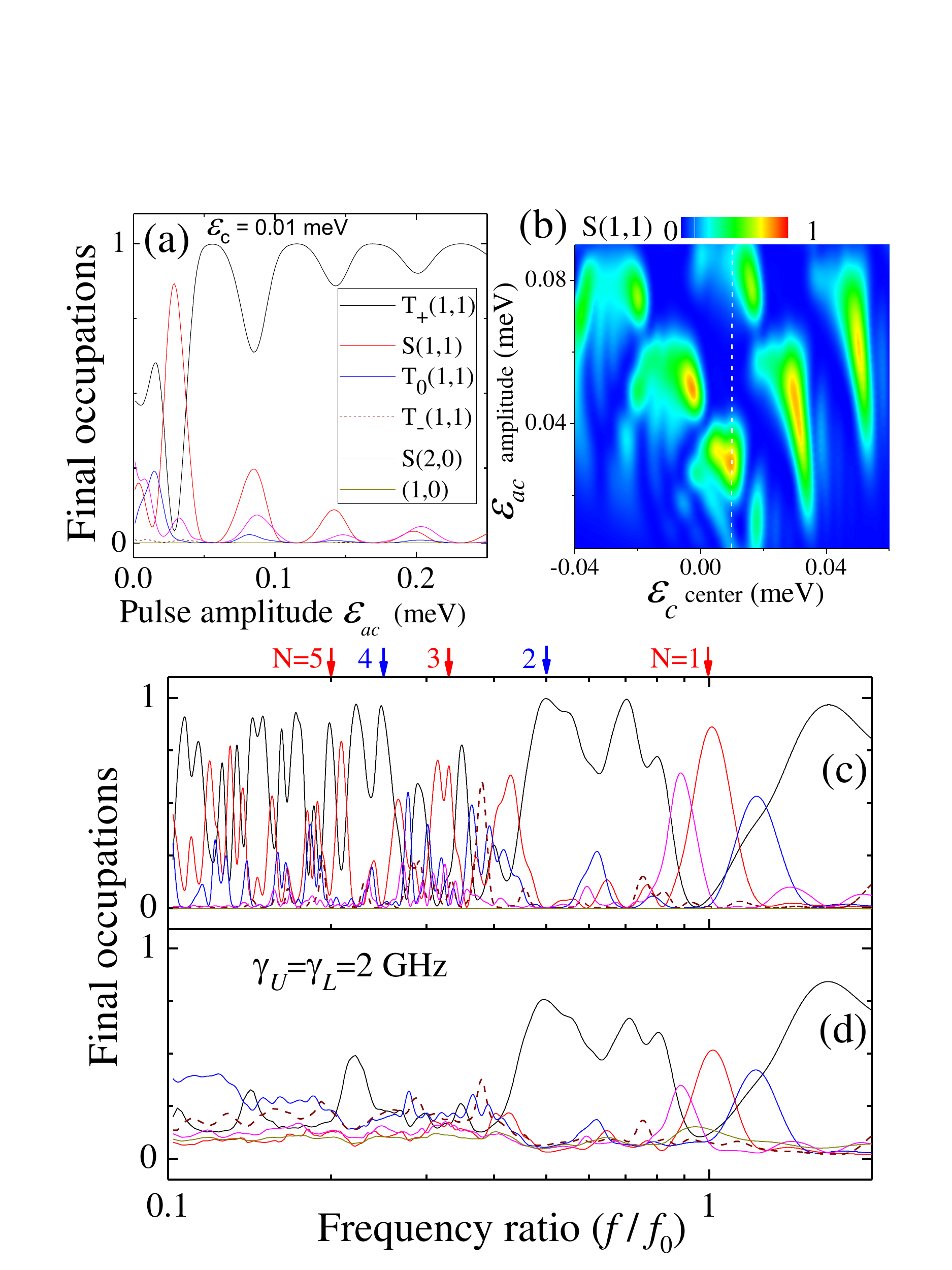} 
\caption{
Results for sinusoidal pulses similar to Fig.~3(f),
$\varepsilon(t)=\varepsilon_c+\varepsilon_{ac} \sin (2\pi f_0 t)$, 
where $h f_0 = E_{S(1,1)} - E_{T_+}$ at $\varepsilon_0$=$-$0.04 meV
($f_0$=4.45 GHz).
(a) State occupations, after 5 oscillations, as function of  the detuning pulse amplitude $\varepsilon_{ac}$
and with $\varepsilon_{c}$=0.01 meV.
(b) S(1,1) occupation after five oscillations of the sinusoidal pulse 
 as function of $\varepsilon_{ac}$ and $\varepsilon_{c}$. The vertical dashed line corresponds
 to the scan given in (a).
(c) Final occupations after five oscillations of the sinusoidal pulse as function
of the pulse frequency $f$ in respect to $f_0$ (in $\log$ scale),
with $(\varepsilon_{0},\varepsilon_{c},\varepsilon_{ac})$=($-$0.04, 0.01, 0.03) meV.
(a), (b) and (c) are calculated with zero charge cycle $\gamma_{U(L)}$= 0.
(d) The same as in (c) but with  $\gamma_{U(L)}$= 2 GHz.
The color scheme in (a) applies also to (c) and (d).
Colored arrows indicate some harmonic excitations of the T$_+$(1,1) $\rightarrow$ S(1,1) transition.
} 

\end{figure}
%++++++++++++++++++++++++++++++++++++++++++++++++++++++++++++++++++++++++

In Fig.~4, we plot some results varying the parameters of the sinusoidal pulse. Fig.~4(a) shows the final
occupations (after 20 ns) as a function of the amplitude of the applied pulse $\varepsilon_{ac}$, in which
an oscillatory behavior is observed with a maximum S(1,1) transfer at $\varepsilon_{ac}$= 0.03 meV, 
as used for Fig.~3(e). Figure 4(b) shows the effects of varying both the amplitude $\varepsilon_{ac}$
and the center of oscillation $\varepsilon_{c}$. Again, oscillatory behavior due to interference effects
of the LZ transition can be seen. In particular, very small S(1,1) transfer is achieved if 
$|\varepsilon_{c}| > \varepsilon_{ac}$, that is, if the avoid crossing is not reached during the ac pulse. This
gives a V shaped pattern in Fig.~4(b). Both Figs.~4(a) and 4(b) were obtained for zero
charge cycle, $\gamma_{U(L)}$= 0, in order to enhance the observed effects.
When the charge cycle is included, the occupation transfers decrease. This can be seen
in Figs.~4(c) and 4(d), where the frequency of the ac pulse was varied with respect
to the resonance transition T$_+$(1,1) $\rightarrow$ S(1,1) at $\varepsilon_0$=$-$0.04 meV.
At resonance $f$=$f_0$= 4.45 GHz, a strong S(1,1) transfer is observed. For a little higher frequency 
$f/f_0$= 1.22, similar resonant transfer occurs for T$_+$(1,1) $\rightarrow$ T$_0$(1,1), consistent
with the alignment of the energy levels [Fig.~1(c)]. For  $f/f_0 <$1, a large number of resonance peaks
is seen, which is mostly due to the harmonic excitation of the inter-level transitions. As
measured\cite{petta-prl} and simulated,\cite{rudner, petta-prb} odd (even) harmonics at $\varepsilon_c \simeq$0
enhance (deplete) the inter-level transfer, as pointed by the red (blue) arrows in Fig.~4(c)
for the S(1,1) transfer.
The inclusion of charge transport cycle, Fig.~4(d), retains the effects but in a lesser pronounced way, especially
for lower frequencies, for which longer times of simulations are needed
in order to maintain fixed the 5 oscillations in the pulse,
therefore favoring the action of charge cycle relaxation.

\subsection{Implementation of quantum gates}

\begin{figure}[htbp]
\includegraphics[width=9.0cm]{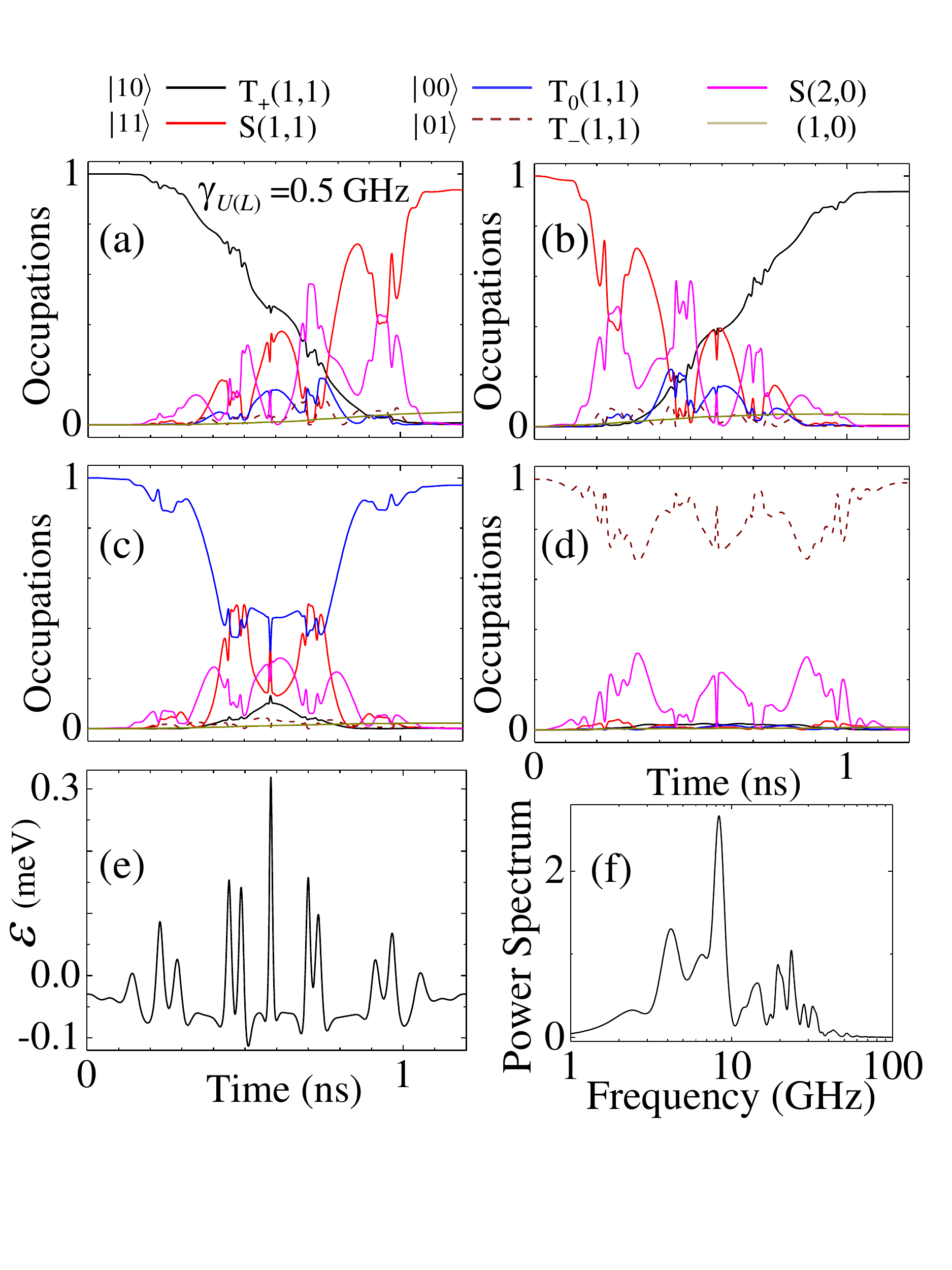} 
\caption{Implementation of the CNOT gate.
Panels (a) to (d): Occupation dynamics for the system initially prepared in states T$_+(1,1)$, S(1,1), T$_0(1,1)$ and T$_-(1,1)$, respectively. (e) Optimized detuning field for the CNOT gate, and (f) corresponding power spectrum, in arbitrary units. The initial and final detuning is $\varepsilon_0$=$-$0.03 meV.}
\label{fig5}
\end{figure}

In order to illustrate the design of a quantum gate, Fig.~\ref{fig5} shows the results of the implementation of the controlled NOT gate in the DQD 
by means of the multi-target control algorithm (cf.~Appendix B).
The CNOT gate operates on two qubits, flipping the second qubit only if the first qubit is in the state $| 1 \rangle  $. In DQD, the qubit basis has been identified as follows: $| 00 \rangle\leftrightarrow{\rm T_0}(1,1)$, $| 01 \rangle\leftrightarrow{\rm T_-}(1,1)$, $| 10 \rangle\leftrightarrow{\rm T_+}(1,1)$, and $| 11 \rangle\leftrightarrow{\rm S}(1,1)$. The charge cycle intensity has been set to $\gamma_{U(L)}$=0.5 GHz, and the multi-target algorithm calculations have been carried out to find the optimized electric field with the final time fixed to 1.2 ns. Panels (a) through (d) show the occupation dynamics for the system initially prepared in states ${\rm T_+}(1,1)$, $S(1,1)$, ${\rm T_0}(1,1)$ and ${\rm T_-}(1,1)$, respectively. It can be noticed that the gate is implemented with relatively high efficiency even under the effects of the very fast decoherent channel, as
also shown in Fig.~6. It should be emphasized that the single optimized electrical field shown in panel (e) is capable of performing the CNOT gate and drives the states $| 00 \rangle \rightarrow | 00 \rangle$, $| 01 \rangle \rightarrow | 01 \rangle$, $| 10 \rangle \rightarrow | 11 \rangle$, and $| 11 \rangle \rightarrow | 10 \rangle$ with fidelity higher than 0.85, as it is shown in Fig.~6.
 The analysis of the Fourier power spectrum of the optimized electrical field, shown in panel (f), reveals its frequency structure, comprising a broad frequencies range up to 70 GHz. Some of the main peaks can be related to state-to-state transition frequencies. For instance, the highest peaks at around 4.2 and 8.3 GHz are close to the resonance frequencies of the
${\rm T_+(1,1)}\rightarrow {\rm S}(1,1)$ and $\rightarrow {\rm T_-}(1,1)$ transitions.
Nevertheless, such optimized field cannot be obtained by adjusting a few number of parameters in a guessed pulse, which emphasizes the relevance of quantum control theory in the implementation of universal quantum gates.

 %++++++++++++++++++++++++++++++++++++++++++++++++++++++++++++++++++++++
% figure-5 
 %++++++++++++++++++++++++++++++++++++++++++++++++++++++++++++++++++++++

\begin{figure}[htbp]
\includegraphics[width=9.0cm]{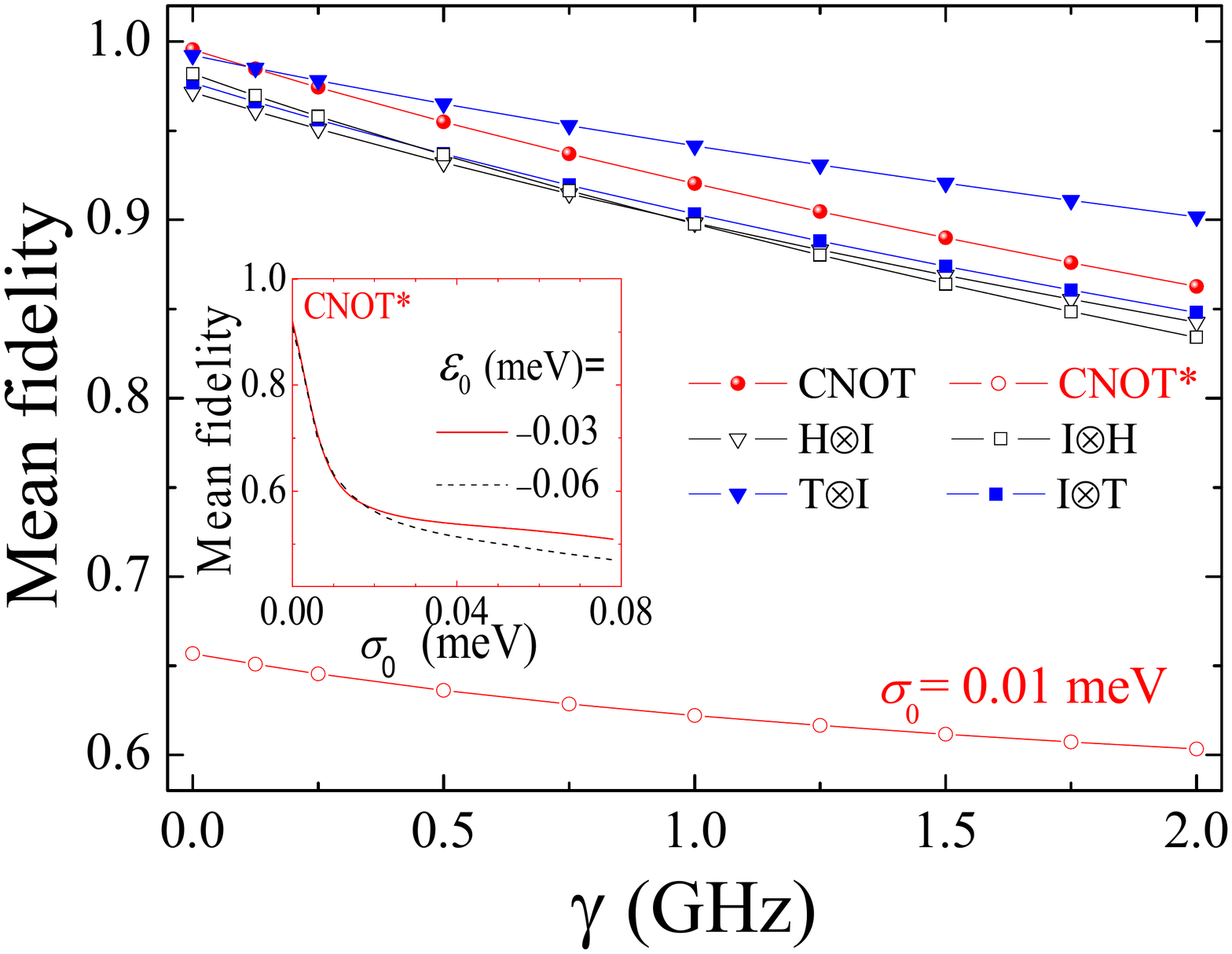} 
\caption{Mean fidelity as a function of the charge cycle intensity 
($\gamma$=$\gamma_L$=$\gamma_U$)
for the universal set of quantum gates: CNOT [red solid (open) spheres without (with) the change noise effect], H$\otimes$I (open triangles), I$\otimes$H (open squares), T$\otimes$I (solid triangles), and I$\otimes$T (solid squares). All cases with detuning $\varepsilon_0$=$-$0.03 meV.
The inset shows the mean fidelity for the CNOT$^*$ gate as a function of the detuning noise spread
$\sigma_0$, for $\varepsilon_0$=$-$0.03 and $-$0.06 meV, and $\gamma$=1 GHz.}
\label{fig6}
\end{figure}

Furthermore, we have made an analysis of the influence of decoherence effects on the following  universal set of quantum gates $\{$CNOT, H$\otimes$I, I$\otimes$H , T$\otimes$I, and T$\otimes$I$\}$, where H is the Hadamard gate, T is the $\pi/8$ gate, and I is the identity. In Fig.~\ref{fig6}, we plot the mean fidelity for these gates as a function of the charge cycle intensity $\gamma=\gamma_L=\gamma_U$. 
An relatively small $\sim$15\% decrease of the mean fidelity is observed when increasing the charge cycle frequency to 2 GHz.
The fidelity $F(\rho,\sigma)=(Tr\left[\sqrt{\sqrt{\rho}\sigma\sqrt{\rho}}\right])^2$ is the measure of the distance between the density matrices $\rho$ and $\sigma$, whereas the mean fidelity of a gate is defined as the mean of fidelities between different initial density matrices $\rho_i$ evolved [like in panels Fig.~\ref{fig5}(a)-(d)] up to the final time $t_f$ and the density matrices obtained from the application of the gate operator $\mathcal{O}$ in the respective initial density matrices, as follows
\begin{eqnarray}
F_m=\dfrac{1}{5}\sum_{j=1}^{5}F\left(\rho_j(t_f),\mathcal{O}\rho_j\right),
\end{eqnarray} 
where  $\rho_1=|00\rangle\langle00|$, $\rho_2=|01\rangle\langle01|$, $\rho_3=|10\rangle\langle10|$, $\rho_4=|11\rangle\langle11|$, $\rho_5=|\psi_5\rangle\langle\psi_5|$, where $|\psi_5\rangle=\left(|00\rangle+|01\rangle+|10\rangle+|11\rangle\right)/2$. As previously discussed, $\rho_5$ is included to prevent relative phase errors
that could occur when dealing separately with $\rho_i$, $i$=1...4.
The pulse duration $t_f$ and the reference detuning $\varepsilon_0$ are parameters that influence the effectiveness of the mean fidelity. The final time $t_f=$1.2 ns and the reference detuning $\varepsilon_0$=$-$0.03 meV maximize the mean fidelity of the CNOT gate and such parameters were used in all results of Fig.~\ref{fig6}.  

In Fig.~7 we plot the optimized detuning fields and the corresponding power spectrum for all one-qubit gates. Through the power spectra, one can notice the frequency decomposition signature of each gate, which has frequencies up to 100 GHz.

 %++++++++++++++++++++++++++++++++++++++++++++++++++++++++++++++++++++++
% figure-6 
 %++++++++++++++++++++++++++++++++++++++++++++++++++++++++++++++++++++++

 %++++++++++++++++++++++++++++++++++++++++++++++++++++++++++++++++++++++
% figure-7 
\begin{figure}[htbp]
\includegraphics[width=9.0cm]{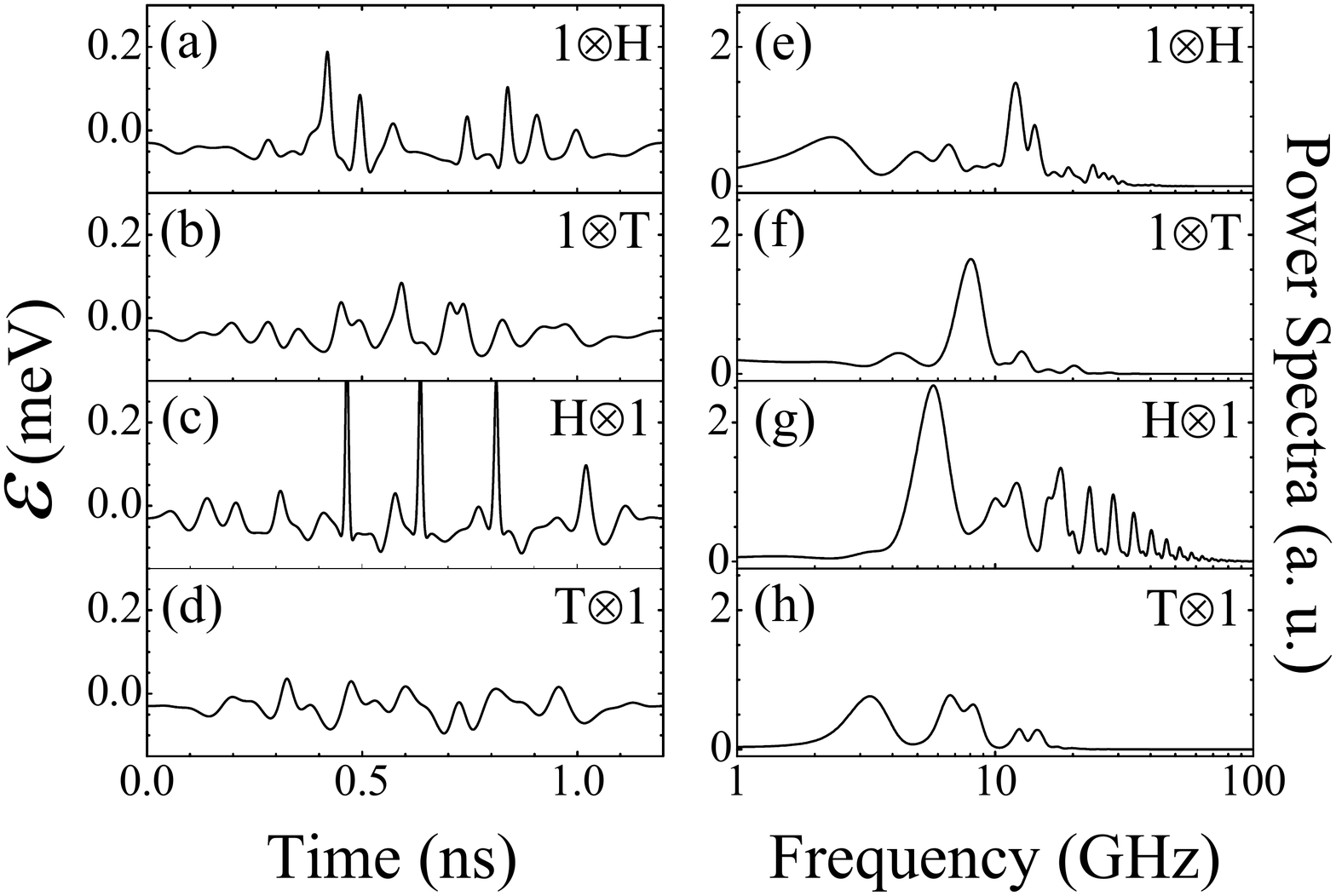} 
\caption{Optimized detuning fields for the following one-qubit gates: I$\otimes$H (a), I$\otimes$T (b),  H$\otimes$I (c), and T$\otimes$I (d). The initial and final detuning is set to $\varepsilon_0$=$-$0.03 meV
 and $\gamma_{U(L)}$=0.5 GHz, as in Fig.~5. The corresponding power spectrum for: I$\otimes$H (e), I$\otimes$T (f),  H$\otimes$I (g), and T$\otimes$I (h).}
\label{fig7}
\end{figure}
 %++++++++++++++++++++++++++++++++++++++++++++++++++++++++++++++++++++++

\subsection{ Theoretical model for the Gaussian noise}

Background charge fluctuations (charge noise) is a significant issue in
experimental conditions in nanowire DQD.\cite{petta-prl}
In order to model this effect in our simulations, we assume the 
charge noise as slow ($<\gamma_{L(U)}^{-1}$) changes in the detuning,
such that the output of many cycle measurements can be seem as
an average over cycles in different detunings.

The dynamics with charge noise can be simulated by the following procedure:
 (i) Evaluate the dynamics given by Eqs.~(\ref{master})-(\ref{m-incoh}) for different
values of reference detuning $\varepsilon'_0$; (ii) Determine the density matrix at the final state by considering an Gaussian average as follows:
\begin{equation}
\rho(\varepsilon_0,t_f) = \int ~ d\varepsilon'_0~ g(\varepsilon'_0,\varepsilon_0)  ~\rho(\varepsilon'_0,t_f), 
\label{master4}
\end{equation}
where
\begin{equation}
g(\epsilon'_0,\varepsilon_0) = \frac{1}{\sigma_0 \sqrt{2 \pi} } \exp{ \left[ 
\frac{-( \varepsilon_0 - \varepsilon'_0 )^2}{2 \sigma_0^2} \right] }, 
\label{av1}
\end{equation}
and $\rho(\varepsilon'_0,t_f)$ is the density matrix at the final time $t_f$, 
calculated for $H (\varepsilon'_0)$. Finally, $\sigma_0$ gives the range of detuning variations. In Fig.~\ref{fig6}, the charge noise effect is presented for the CNOT$^*$ gate as a function of the charge cycle intensity, with $\sigma_0$=0.01 meV.
The inset of Fig.~\ref{fig6} shows the variation of the mean fidelity of the CNOT$^*$ gate 
as a function of the detuning spread $\sigma_0$ for a fixed value of the charge cycle intensity $\gamma=1$ GHz and for two values of initial detuning,  $\varepsilon_0$=$-$0.03 and $-$0.06 meV.
 For up to $\sigma_0 \simeq 0.01$ meV, the mean fidelity decays very fast, afterwards the decay is smoother.
 Although not shown here, we found this to be independent of  $\varepsilon_0$ for values away from the anticrossings, $|\varepsilon_0| \gtrsim 0.02$ meV, where the qubit is usually initialized in a non-mixed
 spin state. Therefore,  $\sigma_0 \simeq 0.01$ meV sets an upper limit for controlling the noise degradation effects in high-fidelity qubit operations.

\section{Conclusions}

We have investigated the control of qubits dynamics in nanowire DQDs.
The eigenstates of the system were solved for two-electron occupation in a quasi-one dimensional
model, including spin-mixing via spin-orbit interaction. The eigenstates were
used to construct a model for the charge transport cycle in the DQD. The transport model 
incorporates the spin mixing and the spatial distribution of charge in the dots.
In this way, only two free parameters ($\gamma_{L(U)}$) are needed, and they control the
intensity of the transport cycle. Aiming at obtaining fast spinflip dynamics, the simulations 
were performed for detunings close to the energy level avoid crossings, 
where charge cycle effects are more important. For simple profiles of the detuning pulse
(stepped and sinusoidal), fast ($\sim$ns) triplet-singlet
transitions are possible with high efficiency as long as the singlet state S(2,0) is occupied
for short times ($\gamma_{L(U)}^{-1}$). The sinusoidal pulse takes advantage of faster and multiple passages,
and of the cumulative effect of LZ transitions, allowing for higher efficiency when charge cycle is 
present.

We have also simulated the set of universal quantum gates in a very fast time scale $\sim$1 ns with a high fidelity. To achieve such gates, we have employed the multi-target formulation 
of QOCT.
Degradation of about 15\% of the fidelity of the gate operations was observed 
when including fast ($>$1 GHz) charge cycle effects, and additional $\sim$20\% degradation
when charge noise effects were taken into account. Nevertheless, the optimized control fields for the set of universal quantum gates have pulse profiles
which can be tested in future experimental investigations. Moreover, the QOCT is a protocol that can be used to implement quantum gates in any platform. If the system dynamics can be manipulated faster than all decoherent channels, such a protocol will be able to implement quantum gates with great success.

\section*{ACKNOWLEDGMENT}

The authors LKC, EFL and MZM acknowledge financial support from 
FAPESP - Brazil.

\appendix
\section{Master Equation - incoherent time evolution}
The incoherent contributions to the master equation Eq.~(\ref{master-incoh}) are written in terms
of Lindblad superoperator $L[A]\rho = A  \rho  A^{\dagger} - \frac{1}{2} 
\left( A^{\dagger}  A  \rho + \rho  A^{\dagger}  A \right)$. The operator $A$ represents
transitions between the one-electron state $| (1,0) \rangle $ and the
two-electron states  $| \psi_{\alpha} \rangle$, $\alpha$= 1,2, ...,5, 
each transition considered to be independent of each other, i.e.~the load process operator is 
given by $A^{L}_{\alpha} = \sqrt{\gamma_L \, l_{\alpha} }\, | \psi_{\alpha} \rangle \langle (1,0) |$ and the 
unload process operator is given by
 $A^{U}_{\alpha} = \sqrt{\gamma_U \, u_{\alpha}} \, | (1,0) \rangle \langle \psi_{\alpha} |$,
 which are defined at the instantaneous detuning using the basis $\{ | \psi_\alpha \rangle \}$.
 $\gamma_{L(U)}$ is a free parameter used to set the intensity of the charge cycle
 and $l_{\alpha} \, (u_{\alpha})$ are the rates shown in Fig.~2(d) [2(e)] for the load (unload) process.

The projections of the incoherent terms of the master equation onto
the reference eigenstate basis set $\{ | \psi_i \rangle \}$, with the Lindblad terms in
the instantaneous set $\{ | \psi_\alpha \rangle \}$, read
\begin{eqnarray}
\mathcal{D} \left[\rho_{m'n'}(t)\right] &=&   \sum_{\alpha=1}^{5} 
\sum_{\alpha',\beta'=1}^{6}    \Bigg\{  
 G_{m'n',\alpha'\beta'} \, \Big[ L[A^{L}_{\alpha}]\rho \, \Big]_{\alpha' \beta'}    \nonumber \\ &&
+  G_{m'n',\alpha'\beta'}
\Big[ L[A^{U}_{\alpha}]\rho \, \Big]_{\alpha' \beta'} \Bigg\}  ,
\label{inc}
\end{eqnarray}
where $G_{m'n',\alpha'\beta'}$ =$ \langle \psi_{m'} | 
\psi_{\alpha'} \rangle \langle \psi_{\beta'} | \psi_{n'} \rangle$  and 
$\Big[ L[A^{L(U)}_{\alpha}]\rho \, \Big]_{\alpha'  \beta'}$=$\langle \psi_{\alpha'} | 
L[A^{L(U)}_{\alpha}]\rho \,   |  \psi_{\beta'} \rangle$. 
Here, primed indices refer to the extended basis sets including the two-electron
eigenstates $\{ | \psi_\alpha \rangle \}$ plus the one-electron state $| (1,0) \rangle $,
i.e.~$\{ | \psi_{\alpha'} \rangle \}=\big \{  \{ | \psi_\alpha \rangle \}, | (1,0) \rangle  \big \} $.
The state $| (1,0) \rangle $ is considered independent of the detuning, orthogonal
to the other states, and shows no coherent dynamics when Eq.~(\ref{master2}) is extended
to include it, i.e.~$H_{m'n'}$=0 if either $m'$ or $n'$=$| (1,0) \rangle $.

Equation (\ref{inc}) can be recast in the form of Eqs.~(\ref{master-incoh2}) and (\ref{m-incoh})
with all $\Gamma_{\alpha' \beta',\gamma' \delta'}$ being zero, except those listed below:
\begin{eqnarray}
\Gamma_{\alpha' \beta',\alpha' \beta'} = - \frac{1}{2} \, \gamma_U
 \left[  u_{\alpha'} +  u_{\beta'}  \right] ,   {\rm for} \, \alpha'  {\rm and} \, \beta' \neq 6.  \\
\Gamma_{\alpha' 6,\alpha' 6}  = - \frac{1}{2} \left[ \gamma_L \, w + \gamma_U
 \,  u_{\alpha'} \right] , \, \, {\rm for} \, \alpha' \neq 6.  \\
w  = \sum_{\alpha=1}^5  l_{\alpha} \, ,  \\
\Gamma_{\alpha' \alpha',66}  = \gamma_L \,   l_{\alpha'}  , \, \, {\rm for} \, \alpha' \neq 6.  \\
\Gamma_{6 \beta', 6 \beta'}  = - \frac{1}{2} \left[ \gamma_L \, w + \gamma_U\, 
 u_{\beta'}  \right] , \, \, {\rm for} \, \beta' \neq 6.  \\
\Gamma_{6 6, \gamma' \gamma'}  = \gamma_U \,  u_{\gamma'}  , \, \, {\rm for} \, \gamma' \neq 6.  \\
\Gamma_{6 6, 6 6}  = - \gamma_L \, w  .
\end{eqnarray}
In these equations, the index 6 refers to the $| (1,0) \rangle $ state.
\section{Multi-target QOCT}\label{AppendixB}
Multi-target QOCT is related to the precise evolution of an initial set of states to a set of target states through the  control field. The well-known variational~\cite{multitarget} and the Krotov~\cite{Kosloff} methods have been employed to perform such a task. In this study, we employ the monotonically convergent algorithm known as two-point boundary-value quantum control paradigm (TBQCP)~\cite{tbqcp} for pursuing such a task.  We refer to this approach as the multi-target QOCT
 and we apply this procedure to our system in order to find the optimized electric field $F_{opt}(t)$ that performs quantum gates.  
 The method starts with the definition of the boundary conditions, which are the set of initial states described by $\{ \rho_j(0) \}$, where $j=1,\dots,N$ and the desired set of observables $\{ O_j(t_f)\}$, at the final time $t_f$. Observables are evolved backwards (from the final time $t_f$ to the initial time $t=0$) through the following equation 
\begin{eqnarray}\label{oper}
&&\frac{\partial O_j^{(n)}(t)}{\partial t}=\frac{1}{i\text{\ensuremath{\hbar}}}\left[H_{0}-\mu F^{(n)}(t),O_j^{(n)}(t)\right],\nonumber\\ &&O_j(t_f) \rightarrow O^{(n)}_j(0).
\end{eqnarray}
where $H_{0}$ is the time independent Hamiltonian of the system and $\mu=$ is the dipole matrix, whose matrix elements are $\mu_{i,j}=-e\langle\psi_i(x_1,x_2)|\left(x_1+x_2\right)|\psi_j(x_1,x_2)\rangle$.  $F^{(n)}(t)$ is the field in the nth iteration of the method. The set of initial states described by the density matrices $\{ \rho_j(0) \}$ are evolved forward with the equation,
\begin{eqnarray}\label{fwdevol}
&&\frac{\partial \rho_j^{(n+1)}(t)}{\partial t}=\frac{1}{i\text{\ensuremath{\hbar}}}\left[H_{0}-\mu F^{(n+1)}(t),\rho_j^{(n+1)}(t)\right],\nonumber\\ &&\rho_j(0) \rightarrow \rho^{(n+1)}_j(t_f).
\end{eqnarray}
where $F^{(n+1)}(t)$ is the (n+1)st iteration field, which is calculated through the following expression
\begin{equation}
F^{(n+1)}(t)=F^{(n)}(t)+ \eta S(t)\sum_{j=1}^N f^{(n+1)}_{j}(t).\label{fieldn}
\end{equation}
In Eq.~(\ref{fieldn}), $\eta$ is a positive constant, $S(t)$ is a positive function, and the field correction is given by
\begin{equation}
f^{(n+1)}_{j}(t) = -\frac{1}{i\hbar}\textrm{Tr}\left\{ \left[ O_j^{(n)}(t), \mu\right]\rho_j^{(n+1)}(t) \right\}\label{fmu}
\end{equation}
Equations~(\ref{oper}-\ref{fmu}) are solved in a self-consistent way, starting with the trial field $F^{(0)}(t)$ and monotonically increasing the value of the desired physical observable $\langle O_j(t_f)\rangle=\textrm{Tr}\left\{\rho_j(t_f)O_j(t_f)\right\}$. As an example, lets consider the CNOT as the quantum gate that must be implemented. In such a case, the initial set of states are  $\{\rho_1(0)=|00\rangle\langle00|,\rho_2(0)=|01\rangle\langle01|,\rho_3(0)=|10\rangle\langle10|,\rho_4(0)=|11\rangle\langle11|, \,\textrm{and}\, \rho_5(0)=|\psi_5\rangle\langle\psi_5|\}$, where $|\psi_5\rangle=\left(|00\rangle+|01\rangle+|10\rangle+|11\rangle\right)/2$. The last state described by $\rho_5$ is included to prevent relative phase errors. The set of target observables is respectively given by $\{O_1(t_f)=|00\rangle\langle00|,O_2(t_f)=|01\rangle\langle01|,O_3(t_f)=|11\rangle\langle11|,O_4(t_f)=|10\rangle\langle10|, \,\textrm{and}\, O_5(t_f)=|\psi_5\rangle\langle\psi_5|\}$. In all numerical calculations, we have used the following parameters: $\eta=0.0005$, $S(t)=\sin^2{\left(\pi\;t/t_f\right)}$, $F^{(0)}(t)=0.035$~V/cm, and run 4000 iteractions of the multi-target QOCT.


\begin{thebibliography}{99}
\bibitem{comment-triplet20} We have solved the problem for 9 states and observed 
the (2,0)-triplet manyfold $\sim$6 meV above in energy from the S(2,0) state.
\end{thebibliography}
\end{document}